\documentclass[pra,twocolumn,floatfix,a4paper,superscriptaddress]{revtex4}
\usepackage{bm,color,graphicx,amsmath,txfonts}

\usepackage[colorlinks, citecolor=blue,linkcolor=blue]{hyperref}

\newcommand{\diag}{{diag}}

\begin{document}
\title{Gaussian geometric discord, entanglement and EPR-steering of two rotational mirrors in a double Laguerre-Gaussian cavity optomechanics in the presence of YIG sphere}

\author{Noura Chabar}
\affiliation{LPTHE-Department of Physics, Faculty of sciences, Ibnou Zohr University, Agadir, Morocco}	
\author{M'bark Amghar}
\affiliation{LPTHE-Department of Physics, Faculty of sciences, Ibnou Zohr University, Agadir, Morocco}	
\author{S. K. Singh}
\affiliation{Process Systems Engineering Centre (PROSPECT), Research Institute of Sustainable Environment (RISE), Faculty of Chemical and Energy Engineering,
University Technology Malaysia, 81310 Johor Bahru, Malaysia}
\author{Mohamed Amazioug} 
\thanks{amazioug@gmail.com}
\affiliation{LPTHE-Department of Physics, Faculty of sciences, Ibnou Zohr University, Agadir, Morocco}

\begin{abstract}	
	
EPR steering is a nonclassical correlation that exhibits properties intermediate to entanglement and Bell nonlocality, providing a valuable resource for quantum communication and computation. In this work, we propose a theoretical scheme to investigate stationary gaussian quantum steering, entanglement and Gaussian geometric discord of two spatially separated rotating mirrors (Rms) in a double-Laguerre-Gaussian cavity (DLGC). Each cavity is derived by Laguerre-Gaussian (LG) beam, and a Yttrium Iron Garnet sphere (YIG) is injected in the intersection of the two cavities. We employ Gaussian quantum steering to characterize the steerability between the Rms. The logarithmic negativity measure is used to quantify the amount of entanglement. We quantify all nonclassical correlations between the Rms by harnessing the Gaussian geometric discord (GGD) measure. Our results indicate that various physical parameters, including the temperature, detuning of the magnon mode frequency, orbital angular momentum (OAM) of the LG cavity modes, the coupling between magnon and cavity mode, the mass of the Rms, each play distinct roles in establishing ${\rm Rm_1-Rm_2}$ entanglement. We characterize the entanglement of the two Rms under ambient temperature (300 K). Stationary entanglement is optimal by adjusting the values of Rms frequency and photon-magnon coupling. We show that the stationary entanglement is fragile under thermal effects. Besides, the GGD demonstrates strong resilience to thermal noise, and this can be further enhanced by increasing the mass of the Rms. Under experimentally accessible parameters and with adjusted ratio between the angular frequencies, we achieve both one-way and two-way steering. Finally, we address the feasibility of our proposal based on the present experiments.\\

	\textbf{Keywords}: Cavity optomechanics; Laguerre-Gaussian-cavity (LGC); Rotating mirrors (Rms); Gaussian steering; Entanglement; Gaussian geometric discord; Yttrium Iron Garnet (YIG).

\end{abstract}

\date{\today}

\maketitle

\section{Introduction}

Entanglement stands as a fundamental cornerstone of quantum physics, distinguishing it from classical mechanics \cite{ESB}. Its significance extends beyond its role in defining this boundary, as it offers a powerful resource for quantum information processing tasks. Optomechanical coupling, achieved through radiation pressure, offers a promising avenue for preparing and controlling the quantum states of mechanical oscillators \cite{OMs}. Radiation pressure can serve as a valuable tool for generating entanglement. Various studies have recently investigated the characterization of entanglement in optomechanical systems \cite{vitali07,abdi15,asjad19,am1,teklu18,prama1,sohail23,Ys,LGH,SKS21,Mekonnen23,shakir18,eleuch,prama2,hmoch}. Moreover, cavity optomechanics has been utilized in a variety of quantum information processing applications, particularly in cooling mechanical modes to their quantum ground states \cite{teutfl}, macroscopic quantum superposition states \cite{abdi16}, achieving steady-state entanglement between mechanical and optical modes \cite{2,k3,p}, electromgnetic induce transparency \cite{EIT,EITamghar}, photon blockade \cite{PB} enhancing precision measurements \cite{2p,3p,4p}, and contributing to gravitational-wave detection \cite{5p}, etc. The cavity optomechanical system operates by utilizing radiation pressure on a vibrating mirror \cite{dvidali,prama1, prama2}. This pressure enables the interaction between the mechanical system and the optical field through the exchange of linear momentum, leading to the entanglement of the light field with the vibrating mirror. Such entanglement can be used for quantum information processing tasks, including teleportation and superdense coding, etc. \cite{7v1,7v2}.\\ 
  
Later, Bhattacharya and Meystre \cite{lg} proposed another type of cavity optomechanical system featuring a rotating mirror; they introduced a novel physical coupling mechanism for investigating the quantum features between cavity modes and macroscopic objects exhibiting rotational characteristics. The system consists of one fixed and the other able to rotate around the cavity axis. When the LG beam reflects within the cavity, the rotating mirror exchanges angular momentum with it, causing the reflected cavity mode photon to carry OAM. As a result, there is entanglement between the LG cavity mode and the rotating mirror. The LG beam, as a typical kind of structured light, possesses a helical wavefront and a doughnut-shaped intensity distribution with a hollow at the beam center \cite{1m,2m}. It is common for the LG beam to have an OAM of $l\hbar$ per photon and a phase of $e^{il\phi}$, where $l$ is the topological charge value and $\phi$ is the azimuthal angle \cite{3m}. The LG beams, due to the exchange of OAM \cite{4m}, can exert torque (radiation torque) on objects, with which it is possible to trap and cool the Rms \cite{29m}. Researchers have realized various phenomena using this system, including ground-state cooling of a rotating mirror \cite{29m}, entanglement between an LG cavity mode and a Rm \cite{30m}, entanglement between vibrational and rotational modes of the same macroscopic mirror \cite{31m}, ground-state cooling of a rotating mirror in a DLGC with an atomic ensemble \cite{32m}, and optomechanical second-order sideband effects in an LG rotational-cavity system \cite{36m}, among others.\\

The approach of improving the entanglement between the photon mode of the cavity field and the phonon mode by incorporating atomic ensembles, ions, polymers, and other materials has been proved by various theoretical and experimental investigations \cite{37m,ion}. Additionally, this approach can achieve bipartite atom-light-mirror entanglement \cite{38m}. However, the frequency of the quantum mode remains constant due to the intrinsic properties of the material chosen. Fortunately, YIG offers a solution to this problem \cite{39m,40m}. YIG spherical crystals can be employed in experiments to overcome the limitations posed by fixed-frequency external ensembles. The magnon mode frequency within a YIG sphere exhibits a high degree of tunability through the adjustment of a bias magnetic field \cite{41m,42m}. Moreover, the spin density of a YIG sphere is several orders of magnitude greater than that of other spin ensembles, such as two-level atomic ensembles. The characteristics of YIG render it suitable for establishing strong coupling with cavity field photons in high-precision cavity optomechanical systems \cite{39m,40m,43m}. Additionally, the Kittel mode within the YIG sphere has unique characteristics, including a low damping rate \cite{damping}, strong coherence, and a long coherence time \cite{46m}, making it applicable to quantum electrodynamics (QED) systems and providing ultra-high flexibility for studying the properties of quantum cavity optomechanical systems. The magnons are quanta of collective spin excitations in materials like YIG \cite{yig}.\\

This work, building on the aforementioned research, aims to investigate the GGD, quantum entanglement, and EPR steering between two Rms. We consider a hybrid DLGC system. Each cavity is composed of a fixed mirror (Fm) and a Rm, which is mounted on a support S and can rotate about the cavity axis z. The YIG sphere is placed at the intersection of the two cavities. Our findings demonstrate that bipartite entanglements in our system can exist and exhibit robustness to thermal noise in the accessible parameter regimes. We thoroughly examine the impact of various physical parameters on GGD, bipartite entanglement, and quantum steering. Additionally, we investigate the influence of magnon-photon coupling on bipartite quantum entanglement. The underlying physical mechanisms responsible for these results are discussed. Such a scheme can be used for one-sided device-independent quantum cryptography \cite{key9,key1}, secure quantum teleportation \cite{key12}, and subchannel discrimination \cite{key13}, due to the asymmetric behavior of the Gaussian quantum steering observed. In this study, we present a novel scheme of entanglement generation between rotating mirrors within a double-rotating optomechanical cavity coupled to an YIG sphere. This is in contrast with recent schemes : Tripartite entanglement in a Laguerre-Gaussian rotational-cavity system with an yttrium iron garnet sphere \cite{bs}. Rotational mirror-mirror entanglement via dissipative atomic reservoir in a double-Laguerre-Gaussian-cavity system \cite{atom}.\\

The remainder of the paper is organized as follows: Section II introduces the model Hamiltonian for the DLGC system in the presence of a YIG sphere. In Section III, we discuss the quantum Langevin equations and their steady-state solutions; we also present the quadrature fluctuation equations for our system Hamiltonian. Section IV, we give the explicit expression of logarithmic negativity, Gaussian quantum steering, and GGD between the Rms modes. In Section V, we analyze the impact of various physical parameters on bipartite entanglement, GGD, and quantum steering. Finally, Section VI summarizes our findings.

\section{Model}

\begin{figure}[h]
   	\centering
   		\includegraphics[width=1.1\linewidth, height=1\linewidth]{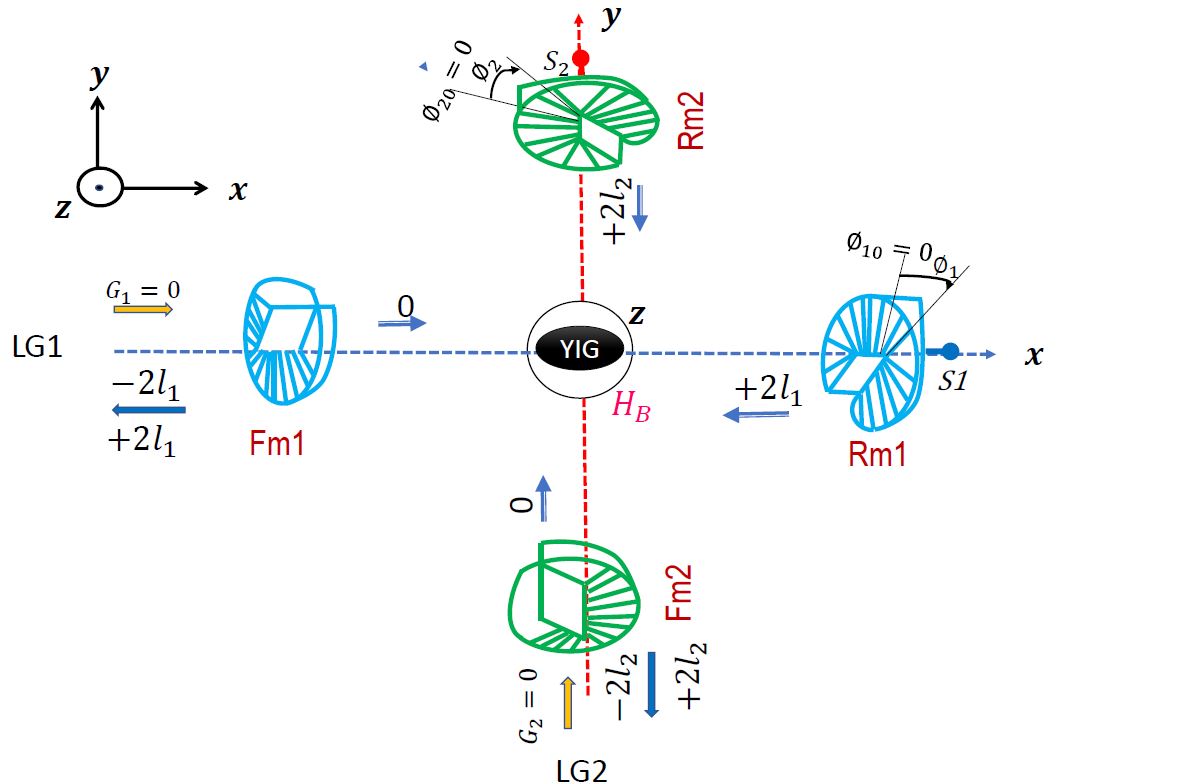}
   	\caption{The DLGC system's schematic diagram: The system features two fixed mirrors (Fm1 and Fm2) and two Rms (Rm1 and Rm2). At the intersection of the two cavities we inject a YIG sphere. The YIG sphere is coupled to the two LG cavity modes through a magnetic field $H_{B}$.  Rm1 and Rm2  are mounted on support points ($S_1$ and $S_2$) and can rotate around the $x$ and $y$ axes, respectively. The fixed mirrors are assumed to be partially transparent, while the Rms are perfectly reflective. The fixed  mirrors enable the Gaussian fields $G_{1,2}$ of optical charge $0$ to penetrate the cavity. The Gaussian beams enter the cavities through the fixed mirrors. The reflected component of the incident beam, which meets the fixed mirrors, is charged with $-2l$, while the transmitted one possesses a charge of zero. The perfectly reflecting mirrors (Rm1 and Rm2) add a charge of $+2l$ to the beam during reflection. The parameter $\phi_{1,2}$ indicates the angular deflection of the two  Rms from equilibrium ($\phi_{10}=0$ and $\phi_{20}=0$). The  charge of the beams at different locations is also indicated. }
   	\label{fig:lg-cavity}
   \end{figure}

The rotational DLGC system consists of two fixed mirrors (Fm1 and Fm2) and two Rms (Rm1 and Rm2), as shown in Fig. \ref{fig:lg-cavity}. The Rm1 and  Rm2 are mounted on the support points $S_{1,2}$ and can rotate around the axes $x$ and z., respectively. We consider that a YIG sphere is placed at the intersection of the two cavities. The fixed mirrors are partially transparent, but they do not alter the topological charge of any beam that passes through them. When the LG beams $G_1$ and $G_{2}$, both with zero charge ($0$), are incident on Fm1 and Fm2, respectively, the reflected component of the incident beam, which meets the fixed mirrors, is charged with $-2l$, while the transmitted one possesses a charge of zero. The two zero-charge beams reflected from Rm1 and Rm2 can acquire a charge of $+2l_1$ and $+2l_2$, respectively. In the equilibrium state, the Rms are at positions $\phi_{10}$, and $\phi_{20}$. When optorotational coupling is influenced, the angular displacements are described by the angles $\phi_{j}$ $(j=1,2)$. Departing from the model described in Ref. \cite{29m}, we postulate the presence of a YIG sphere (a 250-$\mu$ m diameter sphere, as employed in Ref. \cite{45m}) within the two Laguerre-Gaussian cavity optomechanics. The magnon mode frequency within the YIG sphere exhibits a high degree of flexibility in terms of adjustment. The magnon mode within the YIG sphere is excited by the application of a bias magnetic field $H_B$. The coupling between the magnon mode and the LG-cavity mode is achieved through the magnetic dipole interaction. When considering the rotating frame at the driving laser frequency $\omega_{{l}_{j}}$ and applying the rotating-wave approximation (RWA), the Hamiltonian is given by
\begin{equation}
	\begin{aligned}
		H & =  \sum^2_{j=1}\bigg[\hbar \Delta{_{a_{j}}} a_{j}^{\dagger} a_{j}+ 
		\hbar \Delta_{m_j} m^{\dagger} m+\frac{1}{2} \hbar \omega_{\phi_{j}}\left(L_{z_{j}}^2+\phi_{j}^2\right) \\ &
		+\hbar g_{ma_{j}}(a_{j} m^\dagger+a_{j}^\dagger m) 
		-\hbar g_{j} a_{j}^{\dagger} a_{j} \phi_{j}\\ &
		+i \hbar \varepsilon_{l_{j}}\left(a_{j}^{\dagger}-a_{j}\right)\bigg],
	\end{aligned}
\end{equation}
where $\Delta_{a_{j}}=\omega_{a_{j}}-\omega_{l_{j}}$ $ (\Delta_{m_{j}}=\omega_m-\omega_{l_{j}})$ are  the detunings of the cavity photon and the magnon modes respectively with respect to the external driving field with frequencies $\omega_{l_{j}}$.
The operators  $a_{j}$ and $a_{j}^\dagger$ are the annihilation and creation operators of the cavity modes, respectively, with frequencies $\omega_{a_{j}}$, they satisfy the  commutation relation $[a_{j},a_{j}^\dagger]= 1$. Here  $m$ and $m^\dagger$, which satisfy the commutation relation $ [m,m^\dagger]= 1$, represent the annihilation and creation operators of the magnon mode with frequency $\omega_m$. This frequency  $\omega_m$  is determined by  the gyromagnetic ratio $\gamma$ and the  bias magnetic field $H_B$ according to the relation $\omega_m=\gamma H_B$. The quantum operator $L_{z_{j}}$ describes the angular momentum of the $jth$ Rm, while $\phi_{j}$ represents its angular displacement. These operators satisfy the commutation relation $[L_{z_{j}},\phi_{j}]=-i$, and $\omega_{\phi_ j}$ denotes the angular frequency of the jth Rm. The parameters $g_{ma_{j}}$ and $g_{j}$ respectively denote the magnon mode-cavity mode coupling parameters and the optorotational coupling parameters. The optorotational coupling parameters $g_{j}$ are expressed by the relation $g_{j}=cl/L_{j}\left(\sqrt{\hbar/I_{j}\omega_{\phi_{j}}}\right)$, where $I_{j}=m_{j}R^2/2$ represents the moments of inertia of the Rms with mass $m_j$ and radius $R$ about the cavity axis, $c$ is the speed of light, and $L_j$  denotes the lengths of the cavities. The last term represents the Gaussian beam input. The parameters  $\varepsilon_{l_{j}}$ are  related to the input beam powers $P_{L_{j}}$ by the expression $\varepsilon_{l_{j}}=\sqrt{2P_{L_{j}}\gamma_{a_{j}}/\hbar\omega_{l_{j}}}$, and $\gamma_{a_{j}}$ are the cavity damping rates. 
The system's dynamics are described by quantum Langevin equations (QLEs), including noise and damping from Brownian noise and vacuum fluctuations. The QLEs are given by
   \begin{equation}\label{a}
   	\begin{aligned}
   	\partial L_{z_{j}}/\partial t & =-\omega_{\phi_{j}} \phi_{j}-\frac{D_{\phi_{j}}}{I_{j}} L_{z_{j}}+g_{j} a_{j}^{\dagger} a_{j}+\xi_{j}^{\mathrm{in}}, \\
   \partial \phi_j/\partial t & =\omega_{\phi_{j}} L_{z_{j}},\\
   		\partial a_j/\partial t & =-\left(i \Delta_{a_j}+\gamma_{a_{j}}\right) a_{j}-i g_{m a_{j}} m+i g_{j} a_{j}+\varepsilon_{l_{j}}+\sqrt{2 \gamma_a} a_{j}^{\mathrm{in}}, \\
   		\partial m/\partial t & =-\left(i \Delta_{m_{j}}+\gamma_m\right) m-i g_{m a_{j}} a_{j}+\sqrt{2 \gamma_m} m^{\mathrm{in}},
   	\end{aligned}
   \end{equation}
   where $D_{\phi _{j}}$ are the intrinsic damping constants of the rotational mirrors, $\gamma_m$ is the magnon decay rate, and $\xi_{j}^{\mathrm{in}}$ are  the Brownian noise operator which describes the mechanical noise that couples to the rotational mirror from the environment. The operators $a_{j}^{\mathrm{in}}(m^{\mathrm{in}})$ describe the noise operators for the cavities (magnon). The mean values of noise operators $a_{j}^{\mathrm{in}}$ and $m^{\mathrm{in}}$ are zero, and their nonzero correlations  functions are  $\langle a_{j}^{in}(t)a_{j}^{in\dagger}(t^\prime)\rangle=\delta(t-t^\prime)$, and $\langle m^{in}(t)m^{in\dagger}(t^\prime)\rangle=\delta(t-t^\prime)$ \cite{28m}. The mean value of Brownian noise operator $\xi_{j}^{in}$ is zero, and its fluctuations are correlated at the temperature $T$ as \cite{47m}
   \begin{equation}
   	\label{hd}
   	\begin{aligned}
   		\left\langle\delta \xi_{j}^{\mathrm{in}}(t) \delta \xi_{j}^{\mathrm{in}}\left(t^{\prime}\right)\right\rangle= & \frac{D_{\phi_{j}}}{I_{j} \omega_{\phi_{j}}} \int_{-\infty}^{+\infty} \frac{\mathrm{d} \omega}{2 \pi} e^{-i \omega\left(t-t^{\prime}\right)} \omega 
   		\times\left[1+\operatorname{coth}\left(\frac{\hbar \omega}{2 \kappa_B T}\right)\right],
   	\end{aligned}
   \end{equation}
   where $\kappa_B$ is the Boltzmann constant. When the mechanical quality satisfies the condition $I_{j}\omega_{\phi_{j}} /D_{\phi_{j}}\gg1$ Eq. (\ref{hd}) can be reduced to $\left\langle\delta \xi_{j}^{\mathrm{in}}(t) \delta \xi_{j}^{\mathrm{in}}\left(t^{\prime}\right)\right\rangle= \frac{D_{\phi_{j}}}{I_{j} }(2\bar{\mathcal{N}_{j}}+1) \delta(t-t^\prime) $, where  $\bar{\mathcal{N}_{j}}=[\exp(\hbar\omega_{{eff}_{j}}/\kappa_BT-1)]^{-1}$, is the average excitation number at the temperature $T$ and $\omega_{{eff}_{j}}$ are the  effective rotation frequencies  of the rotational mirror. The form of effective rotation frequencies $\omega_{{eff}_{j}}$ satisfies the following relation
   \begin{equation}
   	\label{N}
   	\begin{aligned}
   	\omega_{{eff}_{j}}^{2} & =\omega_{\phi_{j}}^2-\frac{2\xi_{\phi_{j}}^2\gamma_{a_{j}} P_{in_j}}{I_{j}\omega_{a_j}}\left( \frac{\Delta_{a_{j}}}{\Delta_{a_{j}}+  (\gamma_{a_{j}}/2)^2}\right) \times \\  & \frac{[(\gamma_{a_{j}}/2)^2-(\omega^2-\Delta_{a_{j}}^2)]}{[(\gamma_{a_{j}}/2)^2+(\omega-\Delta_{a_{j}})^2][(\gamma_{a_{j}}/2)^2+(\omega +\Delta_{a_{j}})^2]}.
   \end{aligned}
   \end{equation}
   
\section{Covariance matrix of the steady-state}

To examine the dynamics of quantum fluctuations of the operators in Eq. \eqref{a}, we  linearize each operator as the sum of its steady-state value and the fluctuation around that value, and we note  $O=O_s+\delta O$, where $\delta O$ represents the quantum fluctuation, and $O_{s}$ corresponds to  the operators  $a_{j}$, $m$, $L_{{z}_{j}}$,  and $\phi_{j}$ . The steady-state values of the operators in Eq. \eqref{a}   can be obtained in the following form
   \begin{equation}
   	\begin{aligned}
   		a_{s_{j}} & =\frac{\varepsilon_{l_{j}}}{i\Delta_{j}+\gamma_{a_{j}}+\lambda_j}; \quad
   		m_s=-\frac{ig_{ma_{j}}a_{s_{j}}}{i\Delta_{m_{j}}+\gamma_m}; \\  &
   		\phi_{s_{j}}=\frac{g_{{j} |a_{s_{j}}|^2}}{\omega_{\phi_{j}}}; \quad
   		L_{z_{j},s_{j}}=0, 
   	\end{aligned}
   \end{equation}	
   where $\Delta_{j}=\Delta_{a_{j}}-g_{_{j}}\phi_{s_{j}}$,  are the effective cavity detunings, and $\lambda_j=g_{ma_{j}}^2/(i\Delta_{m_{j}}+\gamma_m)$. The steady-state amplitudes $a_{s_{j}}$ can be made real by adjusting the driving field phase, written as
   \begin{equation}
   	a_{s_{j}}=\frac{\varepsilon_{l_{j}}}{\sqrt{\left( \gamma_{a_{j}}+\frac{g_{ma_{j}}^2\gamma_m}{\gamma_m+\Delta_{m_{j}}^2}\right) \left( \Delta_{j}-\frac{g_{ma_{j}}^2\Delta_{m_{j}}}{\gamma_m+\Delta_{m_{j}}^2}\right) }}.
   \end{equation}	
Additionally, we define the quadrature operators for the cavities modes and the magnon mode, along with their respective input noise operators, as follows:  $\delta X_j=\left(\delta a_{j}+\delta a_{j}^{\dagger}\right) / \sqrt{2}, \delta Y_j=\left(\delta a_{j}-\delta a_{j}^{\dagger}\right) / i \sqrt{2}, \delta x=\left(\delta m+\delta m^{\dagger}\right) / \sqrt{2}$,  $\delta y=\left(\delta m-\delta m^{\dagger}\right) / i \sqrt{2}$,  $X^{i n}=\left(a_{j}^{i n}+a_{j}^{i n, \dagger}\right) / \sqrt{2}, Y^{i n}=\left(a_{j}^{i n}-a_{j}^{i n \dagger}\right) / i \sqrt{2}, x^{i n}=\left(m^{i n}+m^{i n \dagger}\right) / \sqrt{2}$,   $y^{i n}=\left(m^{i n}-m^{i n \dagger}\right) / i \sqrt{2} (j=1,2)$. We  obtain the linearized QLEs    
  
   \begin{equation}
   	\label{eq7}
   	\begin{aligned}
   		\partial \delta \phi_{j}/\partial t & =\omega_{\phi _{j}} \delta L_{z_{j} },\\
   		\partial \delta L_{z_{j}}/\partial t & =-\omega_{\phi_{j}} \delta \phi _{j}-\gamma_{\phi _{j}} \delta L_{z_{j}}+G_{{j} } \delta X_{j}+\delta \xi_{j} ^{\mathrm{in}} , \\
   		\partial \delta X_{j}/\partial t & =\Delta_{j} \delta Y_{j}-\gamma_{a_{j}} \delta X_{j}+g_{m a_{j}} \delta y+\sqrt{2 \gamma_{a_{j}}} X_{j}^{\mathrm{in}} , \\
   		\partial\delta Y_{j}/\partial t & =-\Delta_{j}  \delta X_{j}-\gamma_{a_{j}} \delta Y _{j}-g_{m a_{j}} \delta x+G_{{j}} \delta \phi_{j}+\sqrt{2 \gamma_a} Y_{j}^{\mathrm{in}}, \\
   		\partial\delta x/\partial t & =\Delta_m \delta y-\gamma_m \delta x + g_{m a_{j}} \delta Y+\sqrt{2 \gamma_m} x^{\mathrm{in}},  \\
   		\partial\delta y/\partial t & =-\Delta_m \delta x-\gamma_m \delta y-g_{m a_{j}} \delta X+\sqrt{2 \gamma_m} y^{\mathrm{in}},
   	\end{aligned}
   \end{equation}	
   where $G_{{j}}=\sqrt{2}g_{j} a_{s_{j}}$ represents the effective optorotational coupling parameters. For simplicity, Eq. (\ref{eq7}) can be rewritten in matrix form
   
  \begin{equation} \label{QLEs}
   		\partial\mathbf{u}/\partial t=\mathcal{A} \mathbf{u}(t)+\mathbf{n}(t),
   	\end{equation}
   	where  $\mathbf{u}(t) = \begin{pmatrix} \delta \phi _{1}, \delta \phi _{2}, \delta L_{z_{1}},\delta L_{z _{2}},\delta X_{1},\delta X_{2}, \delta Y_{1},\delta Y_{2}, \delta x, \delta y \end{pmatrix}^T$ is the vector of fluctuations, and   $\mathbf{n}(t) = \begin{pmatrix} 0, \xi_{1}^{\text{in}},0 , \xi_{2}^{\text{in}},\sqrt{2\gamma_{a_{1}}} X_{1}^{\text{in}}, \sqrt{2\gamma_{a_{1}}} Y_{1}^{\text{in}}, \sqrt{2\gamma_{a_{2}}} X_{2}^{\text{in}}, \\ \sqrt{2\gamma_{a_{2}}} Y_{2}^{\text{in}},  \sqrt{2\gamma_m} x^{\text{in}}, \sqrt{2\gamma_m} y^{\text{in}} \end{pmatrix}^T$ is the input noise vector, input noises vector with. The drift matrix $\mathcal{A}$ for this system is given by
   		\begin{equation}
  		\mathcal{A}=\left( \begin{array}{cc}
  			\mathcal{q} & B \\
  			N & I
\end{array} \right),		
\end{equation} 
where 
\begin{equation}
	\mathcal{q}= \left( \begin{array}{ccccc}
		0 & \omega_{\phi_{1}} & 0 & 0 & 0  \\
		-\omega_{\phi_{1}} & -\gamma_{\phi_{1}} & 0 & 0 &  G_{{1}} \\
		0 & 0 &  0 & \omega_{\phi_{2}} & 0  \\
		0 & 0 & -\omega_{\phi_{2}} & -\gamma_{\phi_{2}} & 0\\
		0 & 0 & 0 & 0 & -\gamma_{{a_{1}}}
	\end{array} \right),
\end{equation}
\begin{equation}
	N= \left( \begin{array}{ccccc}
		G_{{1}} & 0 & 0 & 0 & -\Delta_1  \\
		0 & 0 &  0 & 0 & 0 \\
		0 & 0 & G_{{2}} & 0 & 0  \\
		0 & 0 & 0 & 0 & 0 \\
		0 & 0 & 0 & 0 & -g_{ma_{1}} 
	\end{array} \right),
\end{equation}
\begin{equation}
	B= \left( \begin{array}{ccccc}
		 0 & 0 & 0 & 0 & 0 \\
		 0 & 0 & 0 & 0 & 0 \\
		 0 & 0 & 0 & 0 & 0 \\
		0 & G_{{2}}& 0 & 0 & 0\\
	 \Delta_1  & 0 & 0 & 0 & g_{ma_{1}}\\
	\end{array} \right),
\end{equation}
and 
\begin{equation}
	I= \left( \begin{array}{ccccc}
			-\gamma_{a_{1}} & 0 & 0 & -g_{ma_{1}} & 0 \\
		0 & -\gamma_{a_{2}}&  \Delta_2 & 0 & g_{ma_{2}} \\
		 0 & -\Delta_2 & -\gamma_{a_{2}} & -g_{ma_{2}}&0 \\
	 g_{ma_{1}}  & 0 & g_{ma_{2}} & -\gamma_m & \Delta_m\\
		 0 & -g_{ma_{2}} & 0 & -\Delta_m & -\gamma_m
	\end{array} \right).
\end{equation}
According to the Routh–Hurwitz criterion \cite{chwartz}, the system is stable. The state of the system under consideration, can be described in the stationary regime using $10 \times 10$ covariance matrix (CM) $\mathcal{Y}$, its elements writes as 
\begin{equation} \label{CM}
\mathcal{Y}_{ij}= \langle\mathbf{u}_i (\infty)\mathbf{u}_j (\infty) +
\mathbf{u}_j (\infty)\mathbf{u}_i (\infty)\rangle/2
\end{equation}
Integrating Eq.~(\ref{QLEs}) 
\begin{equation}
\label{QLEs2} 
\mathbf{u}(t)=R(t)\mathbf{u}(0)+\int_{0}^{t}ds R(r)\mathbf{n}(t-r),
\end{equation}
with $R(r)=\exp(\mathcal{A}r)$. If the stability of matrix $\mathcal{A}$ is satisfied, $R(\infty)=0$, the Eq. (\ref{QLEs2}) writes as
\begin{equation}
\label{eq:uinf}
\mathbf{u}(\infty)=\displaystyle\lim_{t\to\infty} \int_{0}^{t}dr \,R(r)\mathbf{n}(t-r),
\end{equation}
The covariance matrix $\mathcal{Y}_{ij}$ which given by Eq. (\ref{CM}), is written as
\begin{equation}
\label{eq:corrmat}
\mathcal{Y}_{ij}=\sum_{k,k'}Y_{ij},
\end{equation}
where $Y_{ij} = \int_{0}^{\infty} \int_{0}^{\infty}drdr'R_{ik}(r)R_{jk'}(r')\psi_{kk'}(r-r')$ and $\psi_{kk'}(r-r')=\langle \mathbf{n}_{k}(r)\mathbf{n}_{k'}(r')+ \mathbf{n}_{k'}(r)\mathbf{n}_{k}(r')\rangle/2$. For a large mechanical quality factor ($Q_{\phi}\gg 1$), thus 
$$\psi_{kk'}(r-r')=\mathcal{D}_{kk'}\delta(r-r')$$
where $\mathcal{D}_{kk'}$ is the diffusion matrix, describes the stationary noise correlations. The  matrix $\mathcal{D}_{kk'}$ is defined by $ D_{kk'} \delta\left(t-t^{\prime}\right)=\left\langle \mathbf{n}(t)_k(t) \mathbf{n}(t)_{k'}\left(t^{\prime}\right)+\mathbf{n}(t)_{k'}\left(t^{\prime}\right) \mathbf{n}(t)_k(t)\right\rangle / 2$, determined as $\mathcal{D}=\operatorname{diag}\left[0, \gamma_{\phi_{1}}(2 \bar{n}_{1}+1)\right.$, 0, $\gamma_{\phi_{2}}(2 \bar{n}_{2}+1)$ , $\left.\gamma_{a_{1}}, \gamma_{a_{1}}, \gamma_{a_{2}},\gamma_{a_{2}},\gamma_m, \gamma_m\right]$.
Then the Eq.~(\ref{eq:corrmat}) can be writes as
\begin{equation}
\label{eq:Cint}
\mathcal{Y}=\int_{0}^{\infty}drR(r)\mathcal{D}R(r)^{T},
\end{equation}
through integration by parts, the Lyapunov equation for the steady-state covariance matrix, denoted by $\mathcal{Y}$, can be derived as 
\begin{equation}
	\label{1n}
	\mathcal{A} \mathcal{Y}+\mathcal{Y} \mathcal{A}^T+\mathcal{D}=0.
\end{equation}
From Eq. (\ref{1n}), the covariance matrix $\mathcal{Y} $ can be expressed in the form of a block matrix
\begin{equation}
 \mathcal{Y}=\left(\begin{array}{ccccc}
		 \mathcal{Y}_{\phi_{1}} &\mathcal{Y}_{\phi_{1}\phi_{2}} & \mathcal{Y}_{\phi_{1}m} & \mathcal{Y}_{ \phi_{1} a_{1} } & \mathcal{Y}_{\phi_{1} a_{2}} 
		 
		 \\ \\
		\mathcal{Y}_{\phi_{1}\phi_{2}}^T &  \mathcal{Y}_{\phi_{2}} & \mathcal{Y}_{\phi_{2}m} & \mathcal{Y}_{\phi_{2} a_{1}} & \mathcal{Y}_{\phi_{2} a_{2}} \\\\ 
		\mathcal{Y}_{\phi_{1}m}^T & \mathcal{Y}_{\phi_{2}m}^T & \mathcal{Y}_{m} & \mathcal{Y}_{m a_{1}} & \mathcal{Y}_{m a_{2}} \\\\

	\mathcal{Y}_{\phi_{1} a_{1}}^T & \mathcal{Y}_{\phi_{2} a_{1}}^T & \mathcal{Y}_{m a_{1}}^T & \mathcal{Y}_{a_{1} } & \mathcal{Y}_{a_{1} a_{2}}\\\\

	\mathcal{Y}_{\phi_{1} a_{2}}^T & \mathcal{Y}_{\phi_{2} a_{1}}^T & \mathcal{Y}_{m a_{2}}^T & \mathcal{Y}_{a_{1}a_{2} } ^{T} & \mathcal{Y}_{ a_{2}}
	\end{array}\right),
\end{equation}
where each block represents $2 \times 2$ matrix. The diagonal blocks represent the variance within each subsystem (the two Rms modes indicating here by $\phi_1$ and $\phi_2$, cavity mode $a_1$, cavity mode $a_2$, and the magnon  mode $m$. The off-diagonal blocks, on the other hand, represent the covariance between different subsystems, indicating the correlations between two components of the entire coupled DLGC system. 

We evaluate the non-classical correlations in the bipartite subsystem composed of two  Rms  using logarithmic negativity, Gaussian quantum steering, and the GGD. The global CM of the two Rms modes can be represented by the following matrix
\begin{equation}
	\label{covar}
	\mathcal{Y}_{M1M2}=\left[\begin{array}{cc}
	V & \Theta \\
			\Theta^T & F
	
	\end{array}\right],
\end{equation}
the covariance matrices $V=\mathcal{Y}_{\phi_{1}}$ and $F=\mathcal{Y}_{\phi_{2}}$, each with dimensions $2 \times 2$, representing individual modes. The $2 \times 2$ CM $\Theta=\mathcal{Y}_{\phi_{1} \phi_{2}}$ characterizes the correlation between the two Rms. 

\section{Quantum correlations} 
\subsection{Gaussian quantum steering}
 
Gaussian quantum steering refers to the observed asymmetry between two entangled systems, whereby one party can influence or "steer" the state of a remote party by exploiting the shared entanglement. This characteristic can be employed as a measure of the steerability between the two magnons. We employ the covariance matrix (CM) as defined in Eq. (\ref{covar}) and express the Gaussian measurement on Rm1 as in Ref. \cite{steer}:
\begin{equation}
	S_{M_{1} \rightarrow M_{2}}:= \begin{cases}\operatorname{0}, & \text { iff } \mathcal{R} <  1, \\ - \ln  \left(\mathcal{R}\right),  & \text { iff }\mathcal{R} \ge 1,
	 \end{cases}
\end{equation}
where  $\mathcal{R}= \sqrt{\left( \operatorname{det} \mathcal{G}^{M_{2}} \right)} $ is the  simplectic eigenvalues of the matrix $\mathcal{G}^{M_{2}}$ defined as  $\mathcal{G}^{M_{2}}=F-\Theta^T V^{-1} \Theta$, where $V$ , $F$, and $\Theta$, as defined in Eq. (\ref{covar}), constitute the $2 \times 2 $ covariance matrix. The quantification of Gaussian steering from Rm1 to Rm2 is achieved by the function $S_{{M_{1} \rightarrow M_{2}}}$, while the Gaussian quantum steering from Rm2 to Rm1 is determined by the function $S_{{M_{2} \rightarrow M_{1}}}$.
  \begin{equation}
  	\begin{aligned}
  S_{M_{1} \rightarrow M_{2}}:  = \begin{cases}\operatorname{0}, & \text { iff } X <  1  \\  \sqrt{\ln  \left(X\right)} & \text { iff }X \ge 1 \end{cases}  \quad  \text{and}  \\   S_{M_{2} \rightarrow M_{1}}:= \begin{cases}\operatorname{0},  & \text { iff } X <  1   \\ \sqrt{\ln  \left(Y\right)}, & \text { iff }Y  \ge 1 \end{cases},
  \end{aligned}  	
  \end{equation}
  with $X= \sqrt{\ln  \left(\frac{\operatorname{det}(V)}{4 \operatorname{det}(\mathcal{Y}_{M1M2})}\right)} $, and $Y=\sqrt{\ln  \left(\frac{\operatorname{det}(F)}{4 \operatorname{det}(\mathcal{Y}_{M1M2})}\right)} $.  There are three  modes of steerability. The first is "no-way steering," which occurs when $S_ {M_{1}\rightarrow M_{2}} = S_ {M_{2}\rightarrow M_{1}} = 0$, indicating that Rm1 cannot steer the Rm2, and vice versa. The second is "two-way steering", which occurs when $S_ {M_{1}\rightarrow M_{2}} = S_ {M_{2}\rightarrow M_{1}} \textgreater 0$, indicating that the  Rm1 can steer the Rm2 and vice versa. The third is "one-way steering" if  only one Gaussian  $M_{1}$  $\rightarrow $ $M_{2}$ or Gaussian $M_{2}$  $\rightarrow$ $M_{1}$ is steerable. We use the Gaussian geometric discord (GGD) to measure the quantum correlations when the bipartite subsystems Rm1-Rm2 are separable.
  
\subsection{Quantum entanglement}

To quantify quantum entanglement between the two Rms, we use the logarithmic negativity $E_{M_1M_2}$ written as \cite{en0,en1,en2}

\begin{equation}
	E_{M_1M_2}:= \begin{cases}\operatorname{0}, & \text { iff} \quad 2 \xi^{-} <  1,  \\  - ln(2 \hat{\xi^{-}}), & \text {iff} \quad  2\hat{\xi^{-}} \ge 1, \end{cases}
\end{equation}
where $\hat{\xi^{-}}$ is the smallest symplectic eigenvalue measuring the entanglement between the two magnon modes, given by 	
\begin{equation}
	\hat{\xi^{-}}=\left( \frac{ \mathcal{u}-\sqrt{ \mathcal{u}^{2}- \operatorname{det} \mathcal{Y}_{M1M2}}}{2}\right)^{1/2} ,
	\label{ji} 
\end{equation} 
 where, $ \mathcal{u}=\operatorname{det V}+\operatorname{detF}-2\operatorname{det \Theta}$. The two Rms are separable if $\hat{\xi^{-}} > \frac{1}{2}$ (i.e., $E_{M_1M_2}=0$). 
 
\subsection{Gaussian geometric discord}

We adopt the geometric measure of quantum discord, as originally introduced and calculated for two-qubit states in \cite{adis1}, to quantify non-classical correlations within two-mode Gaussian states \cite{adi2,adi3, am1}. A geometric measure of quantum discord quantifies non-classical correlations. It's easily calculated for two-qubit states. Adesso et al. \cite{adis2011} generalize GGD to Gaussian states. The GGD is defined as the smallest squared Hilbert-Schmidt distance between a Gaussian state and the closest classical quantum state that can be obtained by performing a local generalized Gaussian positive operator-valued measurement (GPOVM) on only one party. The GGD between the two Gaussian states, specified by their respective covariance matrices $\mathcal{Y}_{\phi_{1}}$ and $\mathcal{Y}_{\phi_{2}}$, is written as \cite{bs}
\begin{equation} 	
	\begin{aligned}
D_G\left(\mathcal{Y}_{M1M2}\right)  = 		\inf _{F}\left\{D\left(\mathcal{Y}_{M1M2}\right)\right\}   ,
	\end{aligned} 
  	\end{equation}  
where
$$	\begin{aligned}
		D\left(\mathcal{Y}_{M1M2}\right) & =  \frac{1}{\sqrt{\operatorname{det}\mathcal{Y}_{M1M2}}}+  \frac{1}{\sqrt{\operatorname{det}\left(F \oplus V\right)}} \\ &
		\frac{2}{\sqrt{\operatorname{det}\left[\mathcal{Y}_{M1M2}+F \oplus V  / 2 \right]}}    ,
	\end{aligned}
$$
then, the explicit expression of GGD by using the standard form of CM $\mathcal{Y}_{M1M2}$, is written as \cite{bs}
\begin{equation}
	\begin{aligned}
D_G\left(\mathcal{Y}_{M1M2}\right)  =\frac{1}{4(\alpha \beta-c^2)}-\frac{9}{\left(2\sqrt{4 \alpha \beta-  3 c^2}+2\sqrt{\alpha \beta}\right)^2} .
	\end{aligned}
\end{equation}
where $V=\diag(\alpha,\alpha)$, $F=\diag(\beta,\beta)$ and $\Theta=\diag(c,-c)$.

\section{Results and Discussion}

In this section, we will study the evolution of quantum correlations under diverse effects, considering experimentally attainable paradigms. We have taken into consideration the parameters for the DLGC that can be easily achieved in the experiments \cite{30m}, for simplicity we consider that two cavities are symmetric: $R=10 \mu \mathrm{m}, l=100, P_l=50 \mathrm{~mW}$, the laser wavelength $\lambda_{1} = \lambda_{2}=810 \mu \mathrm{m}$, the optical finesse $F=0.5 \times 10^4$, the quality factor $Q_\phi=20\times 10^5, L=1 \mathrm{~mm}, \gamma_{ma} / 2 \pi=1 \mathrm{MHz}$, $g_{ma} / 2 \pi=3.2 \mathrm{MHz},  \omega_{\phi_{1}} / 2 \pi$= $\omega_{\phi_{2}} / 2 \pi=\omega_{\phi} / 2 \pi$=10 $ \mathrm{MHz}$, and $\omega_{eff}\approx\omega_\phi$. We begin by evaluating the density plot of  the bipartite entanglement  $E_{M1M2}$ as a function of the detunings  $\Delta_{m}$ and $\Delta_{a}$ of the magnon mode and the cavity modes, respectively, the mass $m$ of the two Rms,  the orbitals angular momentum $l$, the normalized coupling parameter $g_{ma}$ between magnon mode and the cavity modes, and the temperature $T$. Then, we will explore the evolution of steering $S_{M1 \rightarrow M2}$ and $S_{M2 \rightarrow M1}$, $GGD$, and entanglement between the two Rms as a function of temperature $T$, the mass $m$, the ratio $\omega_{{\phi}_{1}}/\omega_{{\phi}_{2}}$, and the cavity mode-magnon mode coupling $g_{ma}$.
\begin{figure}[h]
  \includegraphics[width=0.45\linewidth]{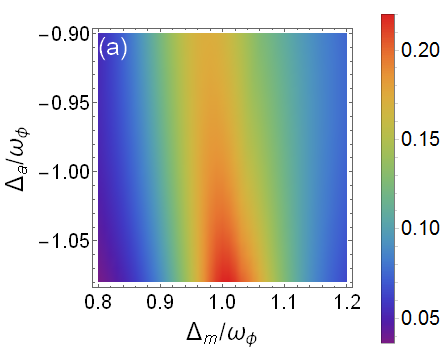}
  \includegraphics[width=0.4\linewidth]{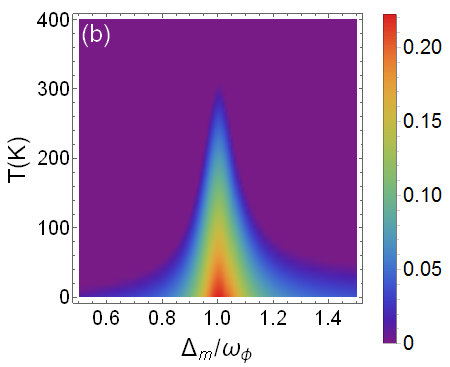}
  \includegraphics[width=0.4\linewidth]{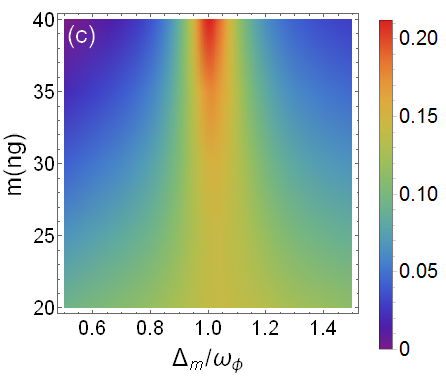}
  \includegraphics[width=0.4\linewidth]{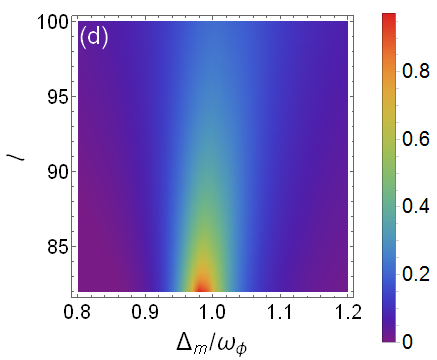}
  	\caption{(a) Density plot of bipartite  entanglement $E_{M1M2}$ as a function of the normalized magnon detuning  $\Delta_{m}/\omega_{{\phi}}$, and the detuning  $\Delta_a/\omega_{{\phi}}$ of the cavity modes. (b) Density plot of bipartite  entanglement $E_{M1M2}$ as a function of the detuning $\Delta_{m}/\omega_{{\phi}}$ and the environmental temperature $T$. (c) Density plot of bipartite  entanglement $E_{M1M2}$ as a function of the normalized magnon detuning  $\Delta_{m}/\omega_{{\phi}}$, and the mass $m$ of the Rms. (d) Density plot of bipartite  entanglement $E_{M1M2}$ as a function of the normalized magnon detuning  $\Delta_{m}/\omega_{{\phi}}$, and the OAM $l$. With $\Delta_a=-\omega_{{\phi}}$, $T=10$ K and $m=40$ ng.}
  	\label{fig:png1}
  \end{figure}
  
We plot in Fig. \ref{fig:png1}(a), the bipartite logarithmic negativity $E_{M1M2}$ of the Rms as a function of the detunings $\Delta_{m}$ and $\Delta_a$. We observe that $E_{M1M2}$ is maximized around sideband $\Delta_{m} = \omega_{\phi}$, corresponding to the blue sideband regime of the Rms. This behavior is ascribed to the cooling effect on the Rms, which aligns with the anti-Stokes process within the cavity. As a result, the decoherence effects are attuned, leading to the maximization of the Rm1-Rm2 entanglement. The level of entanglement can reach $1.25$, which is a significant improvement compared to \cite{bs}, where it does not exceed $0.93$ under optimal parameter adjustments. Furthermore, an increase in detuning $\Delta_{a}$ leads to a significant reduction in Rm1-Rm entanglement.

The Fig. \ref{fig:png1}(b), shows the evolution of the bipartite entanglement $E_{M1M2}$ as a function of ambient temperature $T$ and the detuning $\Delta_{m}$ of magnon mode.  The shared entanglement between the Rms is maximum when $\Delta_{m}=\omega_{\phi}$, as seen and explained in Fig. \ref{fig:png1}(a), after the bipartite entanglement gradually reduce and eventually tend to zero, when the temperature increases. Nevertheless, we remark that the entanglement remains, $>0$, around ambient temperature 300K. This should be observable in experiments under cryogenic conditions. The decoherent phenomenon occurs due to the rise in temperature, which strengthens the decoherence impact of the cavity mode, magnon mode, and phonon mode. Hence, the quantum effect is diminished, as depicted in \cite{bs}.

The bipartite entanglement $E_{M1M2}$ is displays in the fig \ref{fig:png1} (c), as a function of the mass $m$ of two Rms and the detuning $\Delta_{m}$ of the magnon mode. Its clear  that when the mass $m$  increases, $E_{M1M2}$ inecreases. $E_{M1M2}$ reaches its maximum value when  $\Delta_{m}=\omega_{\phi}$ and for $m= 40$ ng. A rise in $m$ around the cooling regime of Rms, can significantly enhance the bipartite entanglement Rm1-Rm2.

Then, in Fig. \ref{fig:png1}(d), we investigate the density plot of entaglement between the Rms as a function of OAM $l$ and the detuning $\Delta_{m}$ of the magnon. As demonstrated, the function $E_{M1M2}$ reaches its optimum value at approximately $\Delta_{m}/ \omega_{\phi}\approx 0.97$, following which it decreases as the OAM increases. As the OAM carried by the LG cavity modes increases, the OAM exchange between the LG cavity modes and mirrors becomes more robust, resulting in an increase in the optorotational coupling $g_{j}$. For $\Delta_{m}/ \omega_{\phi} \gg 1$ and $\Delta_{m}/ \omega_{\phi} \leqq 1$, the entanglement is zero, which can be attributed to being far from the cooling regime of the Rms. As a result, the interaction transfer between the two Rms decreases under the thermal noise imposed by the environment, leading to a reduction in $E_{M1M2}$.\\
\begin{figure}
		\includegraphics[width=0.8\linewidth, height=0.5\linewidth]{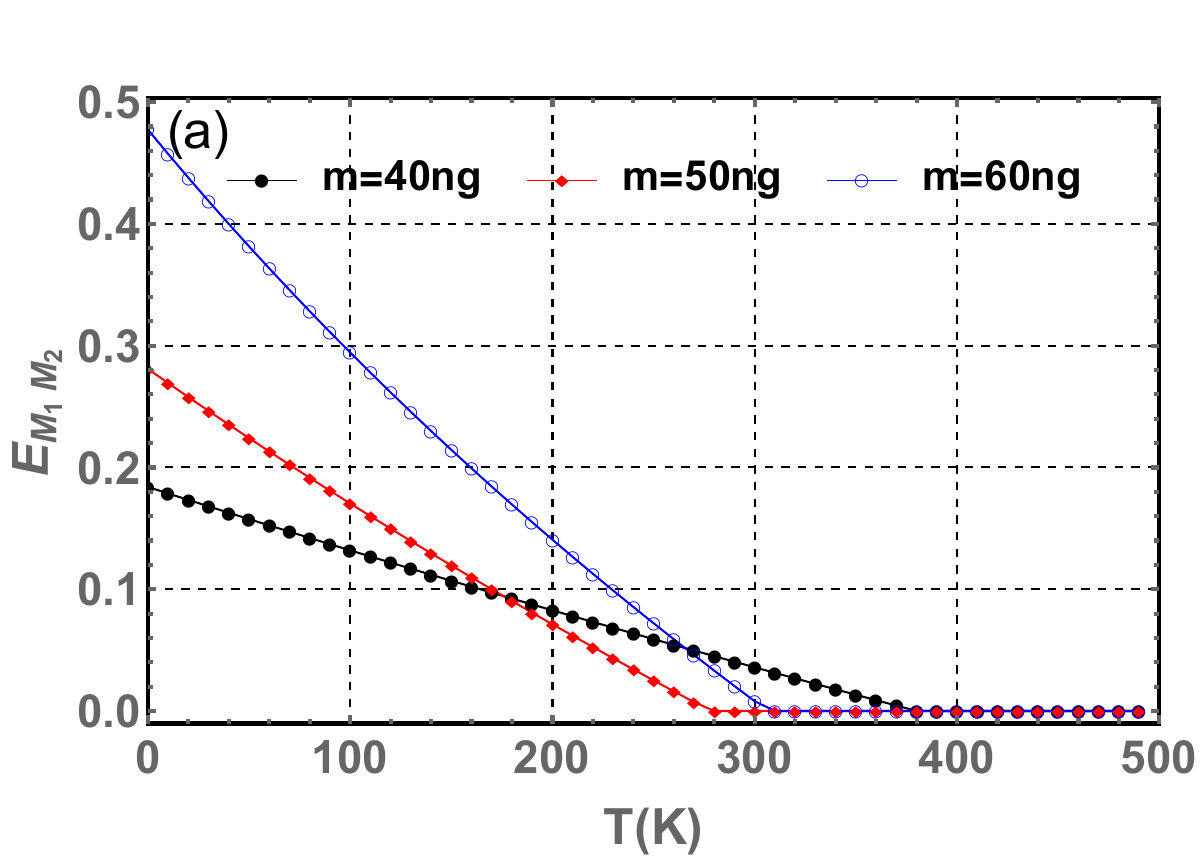}
		\includegraphics[width=0.8\linewidth, height=0.5\linewidth]{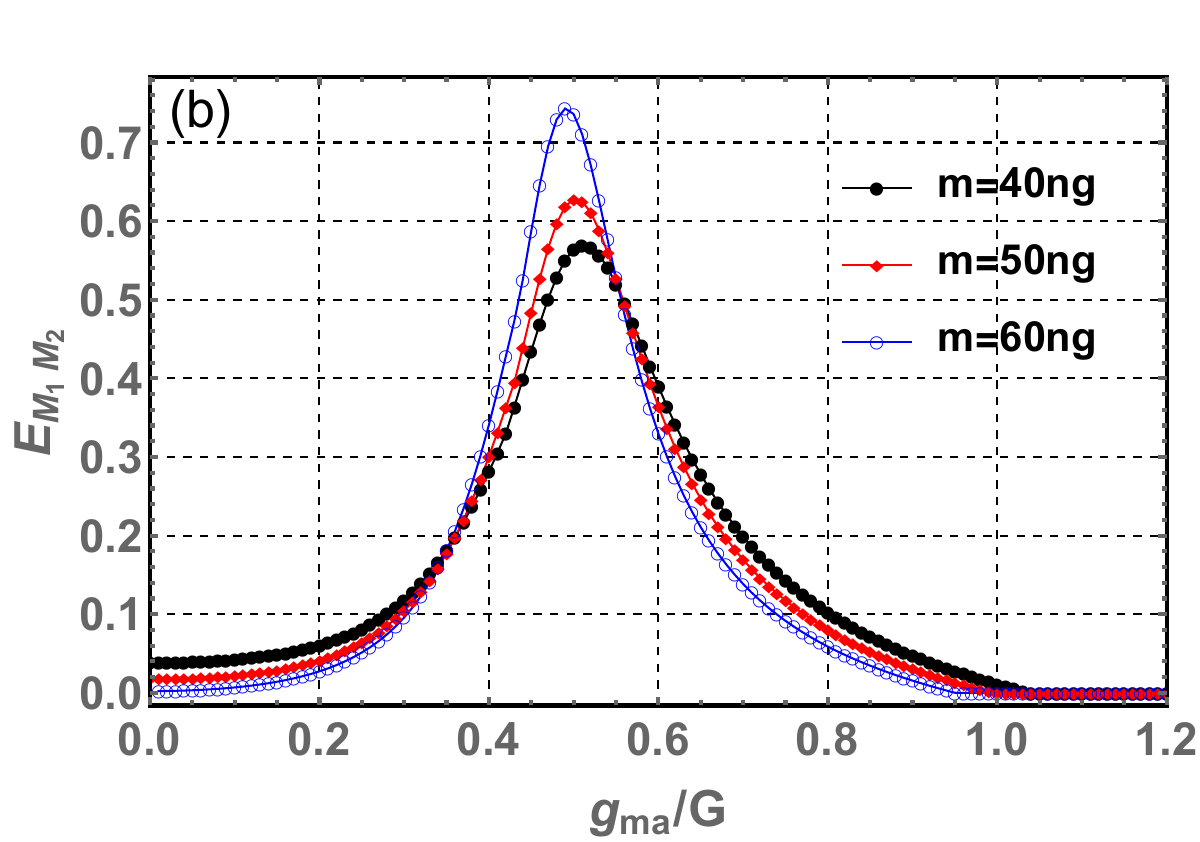}
    	\includegraphics[width=0.8\linewidth, height=0.5\linewidth]{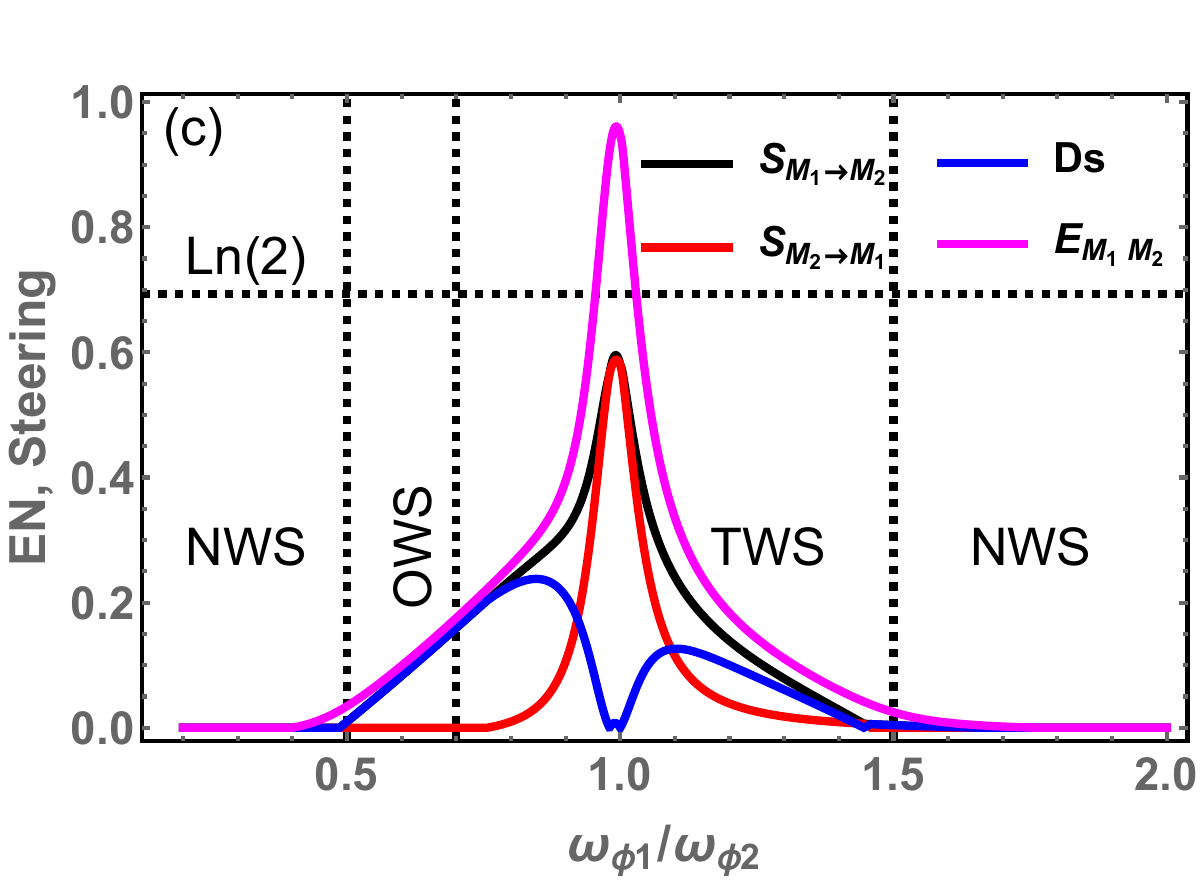}
	   \includegraphics[width=0.8\linewidth, height=0.5\linewidth]{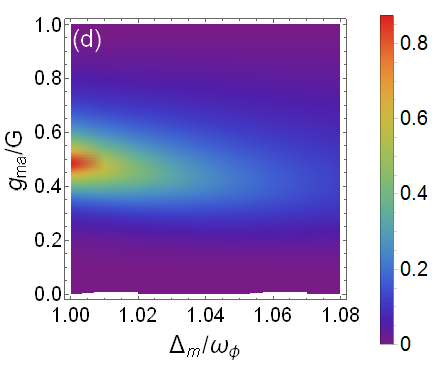}
		
\caption{(a) The logarithmic negativity $E_{M1M2}$ as function of the ambient temperature $T$ with three different mass $m$, (b) The logarithmic negativity $E_{M1M2}$ as a function of the normalized ratio $g_{ma}/G$ with three different
magnon mass $m$ at a fixed temperature $ T = 400$mK. (c) The logarithmic negativity $E_{M1M2}$ as a function of the ratio $\omega_{{\phi}_{1}}/\omega_{{\phi}_{2}}$ at a fixed temperature $T=10$ mK. The other parameters are given in the text. The abbreviations NWS, OWS, and TWS mean no-way steering, one-way steering, and two-way steering, respectively. (d) Density plot of bipartite  entanglement $E_{M1M2}$ as a function of the normalized magnon detuning  $\Delta_{m}/\omega_{{\phi}}$, and the ratio a $g_{ma}/G$.  Here $\Delta_a/\omega_{{\phi}}=-1.07$, $m=60$ng, and  $T=10$ K. The dashed solid horizontal line in (c) describes de value of Ln(2).}
\label{fig:png2}
\end{figure} 
In order to more clarify the relationships between the Rms mass $m$ and the generation of entanglement between the Rms, we plot the logarithmic negativity as a function of the ambient temperature $T$ for various values of the mass $m$ (Fig. \ref{fig:png2}(a)). It is clear that the increase in temperature $T$ contributes to environmental decoherence, which in turn leads to a decrease in the quantum effect. For instance, when $m = 60$ ng, the entanglement between two rotating mirrors drops from its maximum value of \(E_{M1M2} = 0.48\) at $T = 0 $ K to $E_{M1M2} = 0$ at $T = 310 $  K. We define $T = 310$ K as the terminating temperature, denoted by $T_c$. By comparing the four curves in Fig. \ref{fig:png2}(a), it is evident that as the mass increases, the maximum entanglement at \(T = 0 \, \text{K}\) also increases. The terminating temperature \(T_c\) reaches its highest value of about $370 $ K when $m = 40 $ ng. The entanglement for $m=50$ ng disappears quickly and before the two other values of $m$. This phenomenon can be explained by considering the two opposing effects of mass on entanglement: on one hand, a larger mass increases the level of entanglement, while on the other hand, a smaller mass enhances the rotating mirrors robustness against decoherence. The interplay between these opposing effects leads to the observed outcome as mentioned in \cite{effectm}.

In Fig. \ref{fig:png2}(b), we explore the bipartite entanglement between the two Rms as a function of the ratio $g_{ma}/G$ for varying mirror masses. The figure reveals the magnon-photon entanglement transferred to the two spatially separated mirrors. This transfer is mediated by the coupling strength $g_{ma}$: as $g_{ma}$ approaches zero, the entanglement between the Rms ($E_{M1M2}$) vanishes, while for $0 < g_{ma} < G$, we observe $E_{M1M2}> 0$, as shown in Figure \ref{fig:png2}(b). Furthermore, we note that for $g_{ma}<0.5G$, $E_{M1M2}$ increases monotonically before decreasing monotonically to zero at $G \leq g_{ma}$. The maximum achievable entanglement, which occurs around $g_{ma}/G \approx 0.5$, increases with increasing Rms mass. Notably, the shared entanglement between the Rms disappears for a mass of 60ng, earlier than for the other mass cases. The enhancement of entanglement is observed only within a specific range of $g_{ma}/G$ values.\\
In Fig. \ref{fig:png2}(c), we plot the QCs (entanglement $E_{M1M2}$, steering ($S_{M_{1} \rightarrow M_{2}}$, $ S_{M_{2} \rightarrow M_{1}}$), and the Gaussian steering asymmetry ($Ds$)  as a function of the ratio between the angular frequency of the rotating mirrors $\omega_{{\phi}_{j}}$ ($j=1,2$). The first region NWS shows that the entanglement starts to appear when the steering remain null. In the region OWS, we show that there is one-way steering where $S_{M1\rightarrow M2}\neq 0$ and $S_{M2\rightarrow M1}= 0$, i.e., the Rm2 can steer Rm1, but the inverse is not available. In this region, we observe that the steering $S_{M1\rightarrow M2}$, and entanglement are confounded, and both functions increasing with the increases in the ratio $\omega_{{\phi}_{1}}/ \omega_{{\phi}_{2}}$. In the region TWS, we observe that $E_{M1M2}$ takes a higher value than steering, and we observe the generation of steering in the other direction, also, we have witnessed the two-way of quantum steering, which indicate that this region of value of $\omega_{{\phi}_{1}}/ \omega_{{\phi}_{2}}$ is more appropriate to generate QCs between the two mechanical modes. We observe that Gaussian quantum steering remains bounded by entanglement. The steering asymmetry, $Ds$, is consistently less than $\ln(2)$, as depicted in Fig. \ref{fig:png2}(c). It reaches its maximum value when the quantum state is non-steerable in one direction, meaning either $S_{M1\rightarrow M2}>0$ and $S_{M2\rightarrow M1}=0$, or vice versa. As the steerability increases in either direction, $Ds$ decreases \cite{steer}. For $\omega_{{\phi}_{1}}\approx \omega_{{\phi}_{2}}$, both types of steering and entanglement are at their maximum, after they decrease when $\omega_{{\phi}_{1}}/ \omega_{{\phi}_{2}}$ continuously to increase. Eventually, in the last region NWS, all the QCs are no longer present.
 In Fig. \ref{fig:png2}(d), we examine bipartite entanglement $E_{M1M2}$  as a  function of the normalized  magnon-photon coupling and the ratio $\Delta_{m}/ \omega_{\phi}$. We observe that $E_{M1M2}$ remains zero until $g_{ma}/G$ surpasses a certain threshold, at which point it reaches a maximum value before decreasing again as $g_{ma}/G$ increases. While $\Delta_{m}/\omega_{\phi}$ is held at an appropriate fixed value. A thorough analysis of the detuning results indicates that entanglement is maximized when the magnon detuning equals the mirror's effective angular rotation frequency, corresponding to the occurrence of the anti-Stokes process within the system. This condition leads to maximal entanglement between the Rms, underscoring the necessity of cooling the Rms to achieve strong Rm1-Rm2 bipartite entanglement. \\
\begin{figure}[h]
	\centering
	\includegraphics[width=0.8\linewidth, height=0.5\linewidth]{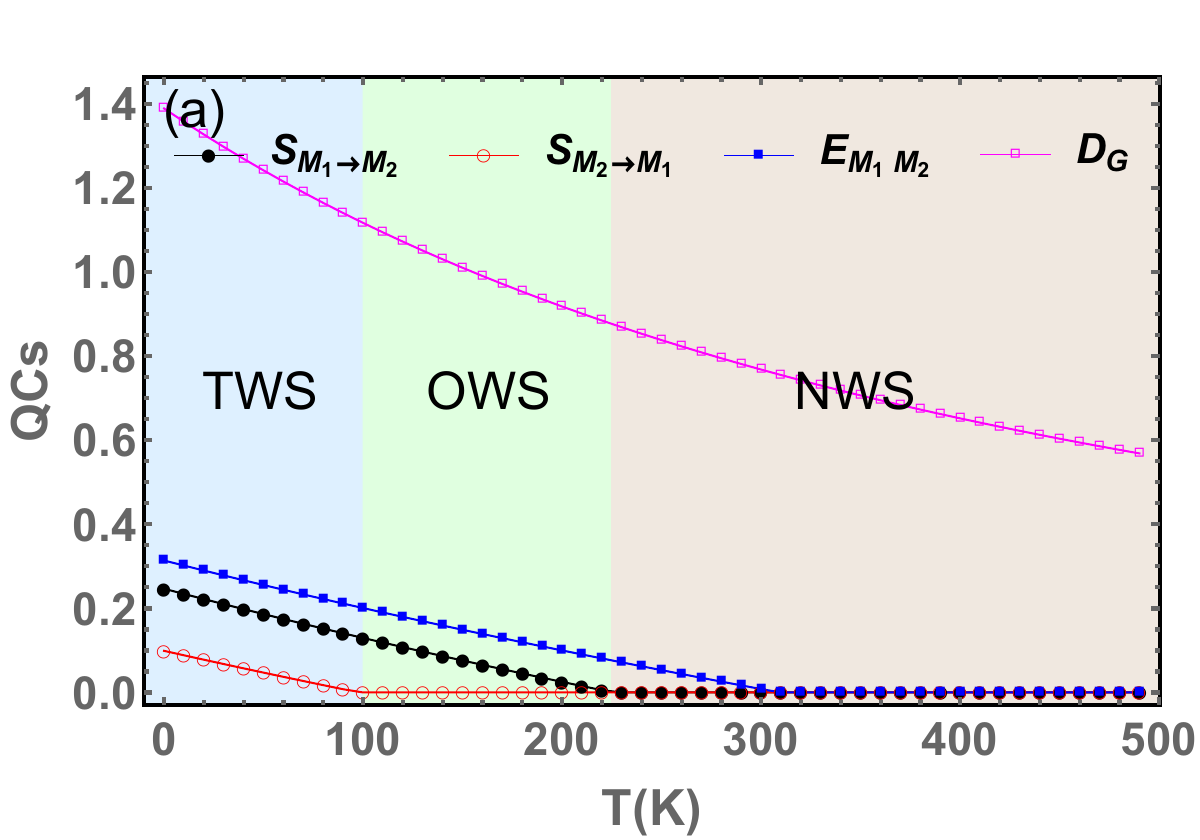}
	\includegraphics[width=0.8\linewidth, height=0.5\linewidth]{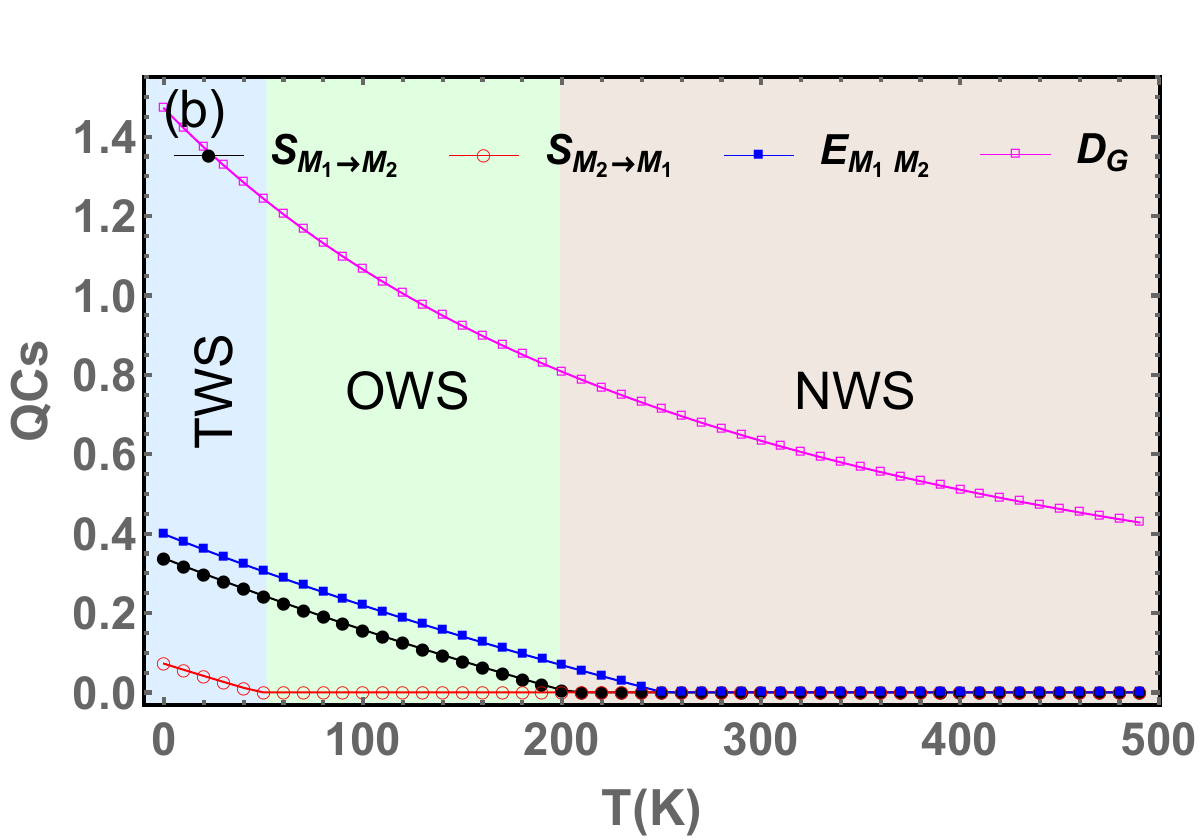}
	\caption{The QCs (entanglement $E_{M1M2}$, steering ($S_{M_{1} \rightarrow M_{2}}$ and $ S_{M_{2} \rightarrow M_{1}}$) and Gaussian geometric discord $D_G$) as function of the ambient  temperature $T$, for $m= 60$ng   (a) and $m= 50$ng (b) and $\omega_{{\phi}_{1}}=0.7\omega_{{\phi}_{2}}$ with $\omega_{{\phi}_{2}}=2\pi\times 10$ MHz.}
	\label{fig:png3}
\end{figure} 
The Fig. \ref{fig:png3} depicts the behavior of the Rm1-Rm2 entanglement in response to temperature variations for two different masses $m$: $m=60$ng (a), and $m=50$ng (b), with $\omega_{{\phi}_{1}}=0.7\omega_{{\phi}_{2}}$. The Qcs decreases with the rise in temperature, due to the temperature-induced  decoherence. It can be clearly seen from Fig. \ref{fig:png3} that GGD is more robust vs thermal noise, and higher than steering and entanglement. The two types of steering are not symmetric, and the amount of steering $S_{M1\rightarrow M2}$ is more important than $S_{M2 \rightarrow M1}$. The logarithmic negativity for $m=60$ng shows greater resistance to thermal effects compared to $m=50$ng, which can be explained by the opposite effect of the mass $m$ aforemontionned  in Fig. \ref{fig:png2}(a). However, the maximum value of entanglement is higher for $m=50$ng different to Figs. \ref{fig:png2}(a) and \ref{fig:png2}(b). There are three regions for steering: TWS, occurring within a temperature range of  $0$ to $90$K for $m=50$ng and between $0$ and $40$K for $m=60$ng, which is considered an adequate range for sharing steering in both directions. After, we pass to OWS and NWS with the increasing of $T$.

\section{conclusion}

In summary, this study theoretically examines the macroscopic entanglement between two Rms within a double-rotating Laguerre-Gaussian cavity system, considering the influence of magnon effects. The LG beams are used to drive the cavities. A YIG is injected in the intersection of the two cavities. The YIG sphere is coupled to the two LG cavity modes through a magnetic field $H_{B}$. We have quantified nonclassical correlations between the Rms modes by considering GGD, entanglement and Gaussian quantum steering. We have examined the evolution of QCS between the Rms of the DLGC system by varying various parameters defined and influencing the system. More importantly, according to the relation examined between the detuning of magnon and the mass $m$, the ambient temperature $T$, the OAM $l$, and the detuning $\Delta_{a}$, the bipartite entanglement can be increased in the blue sideband regime of the Rms. This is consistent with the principles of optomechanical cooling, i.e., the entanglement was maximum at a detuning known to correspond to optimal cooling of the Rms. Under accessible parameters and with a fine-tuned ratio between the angular frequencies, we have achieved both one-way and two-way steering. A rise in $m$ leads to improved entanglement and QCs.  Due to the asymmetric behavior of the Gaussian quantum steering observed, such a scheme can be a useful tool in the context of quantum cryptography, secure quantum teleportation, etc., and a database to control various kinds of nonclassical QCs in macroscopic quantum systems.


\begin{thebibliography}{99}

\bibitem{ESB} 
A. Einstein, B. Podolsky and N. Rosen. Phys. Rev. 47, 777 (1935). E. Schrodinger. Mathematical Proceedings of the Cambridge Philosophical Society 31, 555 (1935). J. S. Bell. On the Einstein Podolsky Rosen paradox. Physics Physique Fizika 1, 195 (1964).
\bibitem{OMs}
T. Corbitt et al., Phys. Rev. A 74, 021802 (2006).
\bibitem{vitali07} D. Vitali, et al., Phys. Rev. Lett. 98, 030405 (2007).
 \bibitem{abdi15} M. Abdi and M. J. Hartmann. New Journal of Physics. 17(1) 013056 (2015).
 \bibitem{asjad19} M. Asjad, N. E. Abari, S. Zippilli and D. Vitali. Optics Express 27, 32427 (2019).
  \bibitem{abdi14xx} M. Abdi, S. Pirandola, P. Tombesi and D. Vitali. Physical Review A. 89, 022331 (2014).
  \bibitem{am1} M. Amazioug, M. Nassik and N. Habiballah. The European Physical Journal D. 72, 1-9 (2018).
  \bibitem{teklu18} B. Teklu. Physics letter A. 432, 128022 (2022).
  \bibitem{prama1}
   M. Amazioug and M.  Daoud. The European Physical Journal. D 75 (6), 178 (2021).
  \bibitem{sohail23}
  A. Sohail, Z. Abbas, R. Ahmed, A. Shahzad, N. Akhtar and J. X. Peng. Annalen der Physik, 535(6), 2300087 (2023).
  \bibitem{Ys}
  A. Hidki, J. X. Peng, S. K. Singh, M. Khalid and M. Asjad. Scientific Reports, 14(1), 11204 (2024).
  \bibitem{LGH}  G. Lai, S. Huang, L. Deng and L. Chen. Photonics. Vol. 10. No. 9. MDPI (2023).
  \bibitem{SKS21}
  S. K. Singh, J. X. Peng, M. Asjad, M. Mazaheri.Journal of Physics B: Atomic, Molecular and Optical Physics. 54(21), 215502 (2021).
  \bibitem{Mekonnen23} H. D. Mekonnen, T. G. Tesfahannes, T. Y. Darge and A. G. Kumela. Scientific Reports. 13, 13800 (2023). 
   \bibitem{shakir18} S. Ullah, H. S. Qureshi and F. Ghafoor. Optics Express, 27(19), 26858-26873 (2019).
	\bibitem{eleuch}
 E.A. Sete, H. Eleuch, C.H.R. Ooi. JOSA B, 31, 2821 (2014).
	\bibitem{prama2}
	N. Chabar,  M. B. Amghar, M. Amazioug and M. Nassik, The European Physical Journal D, 78, 33 (2024).
	\bibitem{hmoch}
  J. Hmouch, M. Amazioug, M. Nassik, Applied Physics B, 129, 151 (2023).
  \bibitem{teutfl}
  J. Teufel, T. Donner, D. Li, J. Harlow, M. Allman,  K. Cicak, A. Sirois, J. Whittaker, K. Lehnert, R. Simmonds, Nature, 475, 359 (2011).
  \bibitem{abdi16} Mehdi Abdi, Peter Degenfeld-Schonburg, Mahdi Sameti, Carlos Navarrete-Benlloch, Michael J. Hartmann. Physical Review Letters. 116(23), 233604 (2016).
  
  \bibitem{2}
   S. Mancini, V. Giovannetti,  D. Vitali, P. Tombesi, Phys. Rev. Lett. 88, 120401 (2002).
  \bibitem{k3}  M. J. Hartmann, M.B. Plenio, Physical review letters,  101, 200503 (2008).
  \bibitem{p} D. Vitali, S. Gigan, A. Ferreira, H.R. Bohm, P. Tombesi,  A. Guerreiro, V. Vedral, A. Zeilinger, M. Aspelmeyer. Physical review letter. 98, 030405 (2007).
  \bibitem{EIT}
  G. S. Agarwal and S. M. Huang. Phys. Rev. A 81, 041803 (2010).
  \bibitem{EITamghar}
  M. Amghar and M. Amazioug. Enhanced transmission and group delay in optomechanics with an optical parametric amplifier. International Journal of Quantum Information, 2024.
  \bibitem{PB}
  M. Amazioug, M. Daoud, S. K. Singh and M. Asjad. Quantum Information Processing, 22(8), 301 (2023).
  \bibitem{2p} Z.-X. Liu, B. Wang, C. Kong,  L.-G. Si, H. Xiong, Y. Wu. Scientific reports, 7, 12521 (2017).
  \bibitem{3p} H. Xiong, L.G. Si, , Y. Wu, Applied Physics Letters, 110, 171102 (2017).???
  \bibitem{4p} H. Xiong, Z.X. Liu, Y. Wu. Opt. Lett. 42, 3630 (2017).
  \bibitem{5p} C.M. Caves. Physical review letters, 45, 75 (1980).
\bibitem{dvidali}
D. Vitali, S. Gigan, A. Ferreira, H. R. Bohm, P. Tombesi, A. Guerreiro, V. Vedral, A. Zeilinger
and M. Aspelmeyer, Physical review letters, 98, 030405 (2007).
   
   \bibitem{7v1}
   M. A. Nielsen and I. L. Chuang, Quantum Computation and Quantum Information (Cambridge
   University Press, Cambridge, 1st ed 2000.
   \bibitem{7v2}
   C.H. Bennett, G. Brassard, S. Popescu, B. Schumacher,
   J.A. Smolin, W.K. Wootters, Physical review letters. 76, 722
   (1996).
		\bibitem{lg}
	M. Bhattacharya and P. Meystre, Physical review letters, 99, 153603 (2007).
	\bibitem{1m}
	L. Allen, S. M. Barnett  and  M. J. Padgett, Optical Angular Momentum (Bristol: Institute of physics publishing, 2016.
	\bibitem{2m}  L. Allen,  M. W. Beijersbergen, R. J. C. Spreeuw,  and J. P. Woerdman, Physical review A, 45, 8185 (1992).
	\bibitem{3m}
	A. M. Yao  and M. J. Padgett,  Advances in optics and photonics, 3, 161 (2011).
	\bibitem{4m}
	H. He, M. E. J.  Friese , N. R. Heckenberg,  and  H. Rubinsztein-Dunlop, Physical review letters, 75, 826( 1995).

	\bibitem{29m}
	M. Bhattacharya and P. Meystre, Physical review letters, 99, 153603 (2007).
	\bibitem{30m}
	 M. Bhattacharya, P. L. Giscard, and P. Meystre,  Physical review A, 
	77, 013827 (2008).
	\bibitem{31m}
     M. Bhattacharya, P. L. Giscard, and P. Meystre,
	 Physical review A, 77, 030303 (2008).
	\bibitem{32m}
	 Y. M. Liu, C. H. Bai, D. Y. Wang, T. Wang, M. H. Zheng, H. F. Wang, A. D. Zhu, and S. Zhang, Optics express, 26, 6143–6157 (2018).
	
   \bibitem{36m}
   S. H. Kazemi and M. Mahmoudi,
   Physica Scripta, 95, 045107 (2020).
   \bibitem{37m}
   Y. H. Ma and L. Zhou, Journal of Applied Physics,
   111, 865–942 (2012).
 \bibitem{ion}
   	D. Garg, A. Biswas, Physical Review A, 100, 053822  (2019).
   \bibitem{38m}
   C. Genes, D. Vitali, and P. Tombesi, Physical Review A, 77, 050307 (2008).
   \bibitem{39m}
   Y. Tabuchi, S. Ishino, T. Ishikawa, R. Yamazaki, K. Usami, and Y. Nakamura, Physical review letters, 113, 083603 (2014).
   \bibitem{40m}
   X. F. Zhang, C. L. Zou, L. Jiang, and H. X. Tang, Physical review letters, 113,
   156401 (2014).
   \bibitem{41m}
   J. Li, S. Y. Zhu, and G. S. Agarwal, Physical review letters, 121, 203601
   (2018).
   \bibitem{42m}
   J. Li, S. Y. Zhu, and G. S. Agarwal, Physical Review A,  99, 021801 (2019).
   \bibitem{43m}
   H. Huebl, C.W. Zollitsch, J. Lotze, F. Hocke, M. Greifenstein, A. Marx, R. Gross, and S. T. B. Goennenwein, Physical Review Letters, 111, 127003 (2013).
   \bibitem{damping}
   X. F. Zhang, C. L. Zou, L. Jiang, and H. X. Tang.
   Sci. Adv 2, e1501286 (2016).
   \bibitem{46m}
   D. K. Zhang, X. M.Wang, T. F. Li, X. Q. Luo,W. D.Wu, F. Nori, and J. Q.
   You, npj Quantum Information, 1, 5014
   (2015).
    \bibitem{yig}
   M. Xiong, M. Wang, G. Q. Zhang, J. Chen. Physical Review A, 107, 033516 (2023).	
\bibitem{key9}
T. Gehring, V. \text{Händchen}, J. Duhme, F. Furrer, T. Franz, C. Pacher, R. F. Werner, and R.
Schnabel, Nat. Commun. 6, 8795 (2015).
\bibitem{key1} C. Branciard, E. G. Cavalcanti, S. P. Walborn, V. Scarani, and H. M. Wiseman, Phys. Rev. A,
85, 10301 (2012).
\bibitem{key12} Q. He, L. Rosales-\text{Zárate}, G. Adesso, and M. D. Reid, Phys. Rev. Lett. 115, 180502 (2015). 
\bibitem{key13} M. D. Reid, Phys. Rev. A 88, 62338 (2013). 
\bibitem{bs}  H. J. Cheng, S. J. Zhou, J. X. Peng,  A Kundu, Li, X.H.,  L. Jin, X. L. Feng. JOSA B. 38, 285-293 (2021).
\bibitem{atom} 
F. Wang, K. Shen and J. Xu, New Journal of Physics, 24, 123044 (2023). 
\bibitem{45m}
X. F. Zhang, C. L. Zou, L. Jiang, and H. X. Tang,
Sci. Adv 2, e1501286 (2016).
  
  
   
   	\bibitem{28m}
   L. Tian and P. Zoller, Physical review letters, 93, 266403 (2004).
   \bibitem{47m}
   C. Gardiner and P. Zoller, Quantum Noise (Springer, 1991).
 \bibitem{chwartz}
   	E. X. Dejesus and C. Kaufman, Physical Review A, 35 (1987) 5288.
		

	
	
	\bibitem{steer}
	I. Kogias, A. R. Lee, S. Ragy and G. Adesso, Physical Review Letters, 114, 060403 (2015).
	 \bibitem{en0}
	M. B. Plenio, Physical Review Letters, 95, 090503 (2005).
	\bibitem{en1}
	G. Vidal and R. F. Werner, Physical Review A, 65, 032314 (2002).
	\bibitem{en2}
	G. Adesso, A. Serafini, F. Illuminati, Phys. Rev. A 70, 022318 (2004).
	\bibitem{adis1} B. \text { Dakić }, C. Brukner, and V. Vedral, Physical Review Letters, 105, 190502 (2010).
	\bibitem{adi2} P. Giorda and M. G. A. Paris, Physical Review Letters, 105, 020503 (2010).
	
	 \bibitem{adi3} G. Adesso, and A. Datta, Physical Review Letters, 105, 030501 (2010).
	 \bibitem{adis2011} G. Adesso , and D. Girolami,  International Journal of Quantum Information, 9, 1773-1786  (2011).	 
\bibitem{effectm}
	 	Z. Chen, J. X. Peng, J. J. Fu, X. L. Feng, Optics express, 27, 29479-29490 (2019).
	
	\bibitem{JLi} 
	J. Li, S.-Y. Zhu and G. Agarwal. Phys. Rev. Lett. 121, 203601 (2018).
	\bibitem{YShen}
	Y. Shen, G. T. Campbell, B. Hage, H. Zou, B. C. Buchler,
and P. K. Lam, Generation and interferometric analysis of high
charge optical vortices, J. Opt. 15, 044005 (2013).
\bibitem{loworbital}
H. Kaviani, R. Ghobadi, B. Behera, M. Wu, A. Hryciw, S.
Vo, D. Fattal, and P. Barclay, Opt. Express 28, 15482
(2020).



\end{thebibliography}
\end{document}